\documentclass[11pt,twoside]{article}
\usepackage{
  amsfonts,
  amsmath,
  amssymb,
  amsbsy,
  amscd,
  bbm,
  booktabs,
  color,
  enumerate,
  float,
  geometry,
  graphics,
  graphicx,
  latexsym,
  natbib,
  svg,
  times,
  verbatim,
  xcolor,
  multirow}
\usepackage[ruled, vlined]{algorithm2e}

\graphicspath{{figs/}{diagrams/}}

\newcommand{\bX}{{\bf X}}

\newcommand{\bZ}{{\bf Z}}

\newcommand{\bbeta}{{\bf \beta}}

\newcommand{\QED}{\hspace*{\fill}\rule{2.5mm}{2.5mm}}

\newcommand{\boldD}{\boldsymbol{D}}

\newcommand{\boldI}{\boldsymbol{I}}

\newcommand{\boldX}{\boldsymbol{X}}

\newcommand{\boldx}{\boldsymbol{x}}
\newcommand{\boldy}{\boldsymbol{y}}

\newcommand{\indicator}{\mathbbm{1}}

\DeclareMathOperator*{\argmin}{arg\,min}

\newcommand{\Yhat}{\widehat{Y}}

\begin{document}

	\begin{center}
		{\large \bf A flexible forecasting model for production systems}\\
		\vspace{0.1cm}
		{R.\;Hosseini, K.\;Yang, A.\;Chen, S.\;Patra},\\
        {Linkedin Data Science Applied Research Team}\\
	\end{center}

    \begin{abstract}
    This paper discusses desirable properties of forecasting models
    in production systems. It then
    develops a family of models
    which are designed to satisfy these properties:
    highly customizable to capture complex patterns;
    accommodates a large variety of objectives;
    has interpretable components;
    produces robust results; has automatic changepoint detection for trend and seasonality;
    and runs fast -- making it a good choice for reliable and scalable production systems.
    The model allows for seasonality at various time scales, events/holidays,
    and change points in trend and seasonality.
    The volatility is fitted separately
    to maintain flexibility and speed and
    is allowed to be a function of specified features.
\end{abstract}

    \section{Introduction}
\label{sect:introduction}

Forecasting business metrics and quantifying their volatility is of paramount importance
for many industries, including the technology industry.
Long-term forecasts can inform the company executives about expectations
about future growth (e.g.\;daily active users)
or about future resource requirements (e.g.\;server capacity needed).
Short-term forecasts with uncertainty intervals can be used
to detect anomalies in the system,
by comparing the forecasts with observed values.

The area of forecasting time series has a long history
with many models and techniques developed in the past decades.
Some important examples include:
Classical time series models such as ARIMA
(e.g.\;\cite{book-hyndman-2014}) and GARCH (\cite{book-tsay-2010});
Exponential Smoothing Based methods (see \cite{winters-1960});
State-space models (see \cite{kalman-1960}, \cite{book-durbin-2012}, \cite{book-west-1997});
Generalized Linear Models extensions to time series
(\cite{book-kedem-2002}, \cite{hosseini-takemura-2015}); 
Deep Learning based models such as LSTM (\cite{hochreiter-1997}).
Our framework utilizes the powerful aspects of these various models.
We provide more details below, after discussing the motivations for its development.

Here we discuss the desirable properties of a forecasting method in practice,
especially for production systems in the technology industry.
These properties are the motivation for our framework.
\begin{itemize}
    \item Flexibility to accommodate complex patterns for the metric of interest:
    Flexibility is needed to achieve high accuracy.
    For example, in practice, often the growth can differ
    for weekdays versus weekends for a daily metric
    and this needs to be taken into account explicitly or implicitly to achieve accurate forecasts.
    \item Flexibility in the objective: As an example, depending on the application, the user might
    intend to capture the average values well (e.g.\;in Revenue forecasting)
    or the peak values well (e.g.\;in capacity planning).
    \item Interpretability: This can benefit
    users to inspect and validate forecasting models with expert knowledge.
    As an example, if the model is able to decompose the forecast
    into various components (e.g.\;seasonal, long-term growth, holidays),
    the users can inspect those components not only to validate the models
    but also to get insights about the dynamics of the metric.
    \item Robustness: It is important for the forecast values to be robust
    and have low chance of returning values which are implausible.
    This can indeed occur in the time series context for various reasons,
    including the divergence of the simulated values (see e.g.\;\cite{hosseini-takemura-2015}).
    \item Speed: In many applications it is important to quickly train the model and produce forecasts.
    Speed can help with auto-tuning over parameter spaces and producing a
    massive number of forecasts, even when there are many training data points.
\end{itemize}

In our model, we have decomposed the problem into two phases:
\begin{itemize}
    \item Phase (1) the conditional mean model;
    \item Phase (2) the volatility / error model.
\end{itemize}
In (1) a fitted model is utilized to predict the metric of interest
and in (2) a volatility model is fitted to the residuals.
This choice helps us with flexibility and speed as integrated models
are often more susceptible 
to being poorly tractable (convergence issues for parameter estimates), and
demonstrate divergence issues when Monte Carlo Methods are used to generate future values, 
which is the case for many of the aforementioned methods such State-Space Models or GLM based models
(\cite{book-tong-1990}, \cite{hosseini-pcpn-2017}).
As an example, estimating a model with complex conditional mean and complex volatility
(i.e.\; with a large number of parameters)
can run into stability and speed issues (\cite{hosseini-bk-2020}).

Phase (1) can be broken down to these steps:
\begin{itemize}
    \item[(1.a)] extract raw features from timestamps, events data and history of the series;
    \item[(1.b)] transform the features to basis functions;
    \item[(1.c)] apply a change-point detection algorithm to the data to discover changes
    in the trend and seasonality over time;
    \item[(1.d)] apply an appropriate machine learning algorithm to fit those features
    (depending on objective).
\end{itemize}
    Note that, the purpose of Step (1.b) is to transform the features into a space which can used
in ``additive'' models when interpretability is needed.
For Step (1.d), our recommended choices are explicit regularization based algorithms
such as Ridge.
Note that if the objective is to predict peaks,
quantile regression or its regularized versions are desirable choices.
In the next sections,
we provide more details on how these various features are built
to capture various properties of the series.

In Phase (2), a simple conditional variance model can be fitted to the residuals
which allows for the forecast volatility to be a function of specified factors (e.g.\;day of week,
if volatility depends on that).

The main contribution of this work is to develop a model which combines various techniques to 
achieve a highly customizable model which can run fast; have interpretable options and support 
variable objectives. Our events model is very flexible and allows for the deviation due to events
to be of arbitrary shape even within a day. Also the regularization-based
changepoint algorithm is a novel method which can capture changes in both trends and seasonality -- 
we demonstrate it works well in practice.

The paper is organized as follows.
Section \ref{sect:model} discusses the model details.
Section \ref{sect:change-points} provides the details
for the automatic change-point detection component.
Section \ref{sect:use_case} demonstrate how our models work using a particular use case.
In Section \ref{sect:assessment},
we discuss methodology for assessing performance of the 
models in terms of accuracy and in cross-validation.
We also compare the performance of our models 
with some widely used open source libraries.
Finally Section \ref{sect:discussion} concludes with a summary
and discussion of possible extensions.

    \section{Conditional Mean and Volatility Models}
\label{sect:model}
This section discusses the details of our model.
We refer to our model as Silverkite for clarity in the following.

First, we introduce the mathematical notation.
Suppose $\{Y(t)\}, t=0,1,\cdots$ is a real-valued time series where $t$ denotes time.
We denote the available information up to time $t$ by $\mathcal{F}(t)$.
We assume this information is given in terms of a covariate process, denoted by
$\bZ(t)$, as discussed in \cite{book-kedem-2002}.
The covariate process is assumed to be a
multivariate real-valued process and encodes the features to be utilized in the models.
For example, we can consider the covariate process:
\[\bZ(t-1) =  (1, Y(t-1), Y(t-2)),\]
which means the features used in the model are simply a constant and two lags.
(This corresponds to a simple auto-regressive model of order 2, if we further
assume that the conditional mean is linear and the conditional distribution is Gaussian.)

\begin{figure}[H]
    \centering
    \includegraphics[width=1\textwidth]{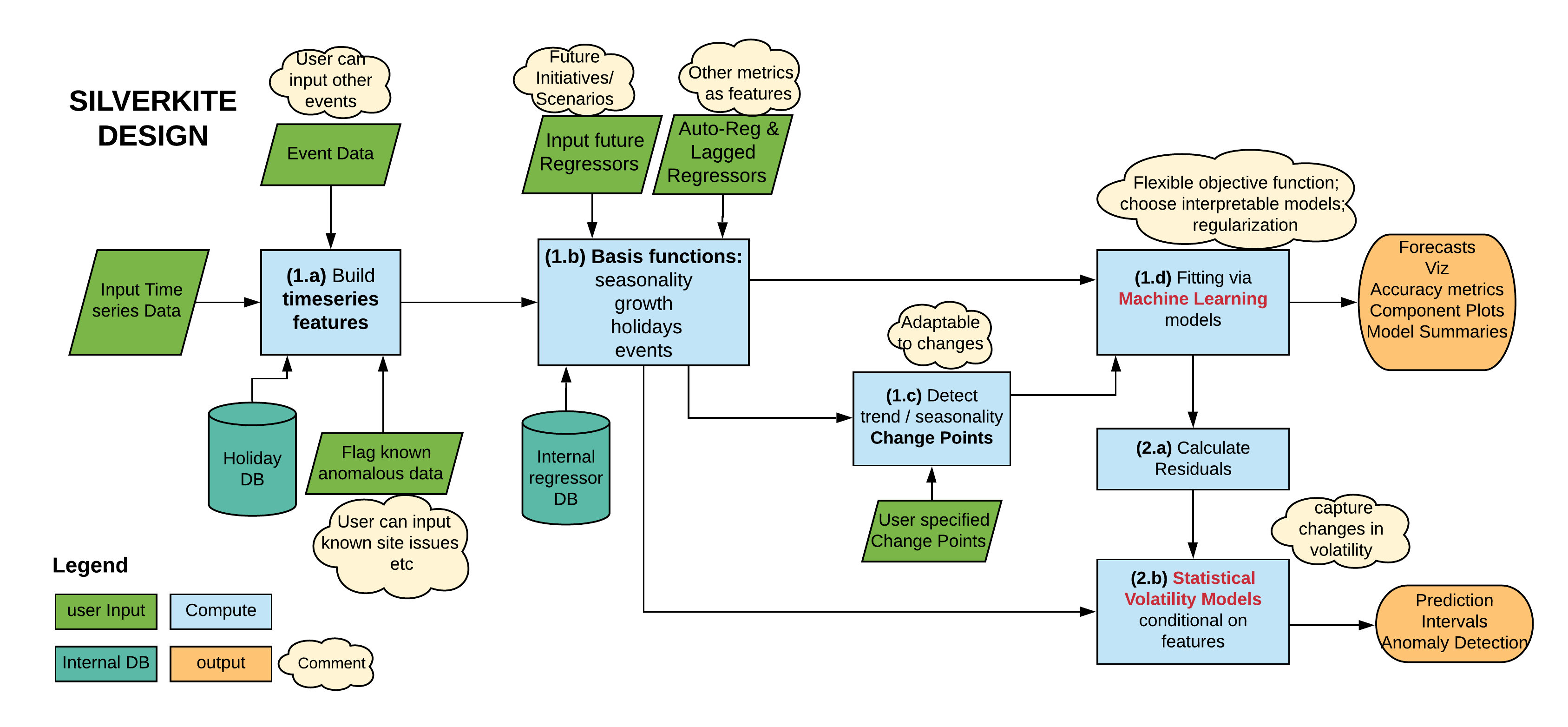}
    \caption{This diagram summarizes the steps involved in the Silverkite Algorithm.
    The square blue notes represents the computational nodes and include a number which represent the step.
    The nodes (1.a) to (1.d) correspond to Phase (1) which is the conditional mean phase.
    The nodes (2.a) and (2.b) correspond to Phase (2) which is the volatility / error model.}
    \label{fig:silverkite-diagram}
\end{figure}

Figure \ref{fig:silverkite-diagram} illustrates the model components diagram where
the (sky) blue square nodes show the compute nodes; the green parallelograms show the
(potential) user inputs; the dark cylinders show the (potential) input databases such as
country holidays; the yellow clouds show the inline comment / descriptions.

As we discussed in the introduction, the algorithm has two main phases.
\begin{itemize}
    \item Phase (1): The conditional mean model
    \item Phase (2): The volatility / error model, fitted to residuals
\end{itemize}
As we discussed, this choice is made deliberately to increase the flexibility of
the model in capturing complex mean patterns / objectives in modeling the mean
and overall speed of the algorithm. The main reason is models which encapsulate 
the mean parameters and volatility parameters into one model often require slow 
Maximum Likelihood Estimation (e.g.\;in \cite{hosseini-takemura-2015})
or intensive Bayesian computation (e.g.\; \cite{book-west-1997})
due to parametric form of the volatility.

\subsection*{The conditional mean component}
\label{subsect:mean-component}

The very first step in building the model is generating the raw features.
Here are some examples of raw features:
\begin{itemize}
    \item Time of day (TOD): This is a continuous measure of what time in the day $t$ is,
        in hour units, and ranges from 0 to 24.
        For example 12.5 is half past noon.
    \item Day of week (DOW):
        Categorical variable denoting day of week, from Sunday (0) to Saturday (7).
    \item Time of week (TOW):
        This is a continuous measure of when in the week $t$ is, in day units,
        starting from Sunday.
        For example, 1.5 denotes Monday at noon.
    \item Time of year (TOY): This is a continuous measure of when in the year $t$ is, in
    year units and ranges from 0 to 1. For example 1.5/365 denotes the second of January at noon
    in a non-leap year.
    \item Time of month (TOM) and time of quarter (TOQ) can be defined similarly and range from 0 to 1.
    \item Event / holiday labels: assume a holiday or re-occurring events database is available.
    For example, timestamps
    on January 1st could be labeled with ``New Year''.
    Custom events can also be treated in the same way.
    If multiple event databases are relevant (e.g.\;Gregorian and Lunar calendar), multiple
    features are to be generated per database.
    \item Continuous Time (CT): This is a variable which measures the time elapsed from
    a particular date in year units. Typically, the reference date is the beginning of the time series,
    so that the first time point has value $0$.
    For example if the series starts on Jan 1st, 2015,
    then end of 2016 will have the value of $2$ since two years have elapsed.
\end{itemize}

\subsubsection*{Seasonality}
\label{subsubsect:seasonality}

The seasonality in a time series can appear in various time-scales.
For example for hourly data, one can expect a periodic daily pattern across the day, a periodic
weekly pattern across the week, and similarly periodic patterns across month, quarter and
year.

We use Fourier series as the primary tool for modeling seasonality in various scales.
For example for hourly data and within day variations, we use Fourier terms of the form:
\[s_k = \sin (k \omega_d d(t)),\;c_k = \cos (k \omega_d d(t)), \;\omega_d = 2 \pi / 24,\]
\[k=1,\cdots,K\]
where $d(t)$ denotes TOD (time of day), discussed above; and $K$ is the appropriate order of 
Fourier series which can be determined in a Model Selection Phase.
Note that the frequency $w_d$ is set to be $2 \pi / 24$ since TOD changes from 0 to 24.
Similarly, we can define appropriate Fourier Series for weekly, monthly, quarterly and annual
periodicity.

\subsubsection*{Growth}
\label{subsubsect:growth}

In order to model growth, we introduce some basis functions and allow for piece-wise
continuous versions of those functions.
By slight abuse of notation let $t$ denote the
continuous time (CT) introduced in the raw features.
Then consider the following basis functions:
\[f(t) = t^p, p = 1/3, 1/2, 1, 2, 3.\]
We allow for the growth to change with time continuously at given change-points
$t_1, \cdots, t_k$.
Given the change-points and a basis function $f(t)$, we define
\[growth(t) = \alpha_0 f(t) + \sum_{i=1}^k \alpha_i 1_{t > t_i} (f(t) - f(t_i))\]
Note that $growth(t)$ is a continuous function of $t$, but allows the derivative of the
function to change at the change-points.

In some forecast applications, we might have external information about the change-points,
which can then be fed to the algorithm.
However, in other applications such information might be unavailable or the growth shape
might not conform to any such basis function.
Therefore Section \ref{sect:change-points} presents a method for automatic change-point detection
both for growth and seasonality.

\subsubsection*{Events and holidays}
\label{subsubsect:events}
Here we present a method for modeling reoccurring events.
This approach allows the period of the event to take any possible deviation 
from general trends. It assumes the impact of the event does not vary over time.

A prime example of such events are national/religious holidays in various countries
and the days surrounding those holidays.

Suppose, the event occurs in the known periods $[t_i, t_i + l]$, where each $i$
corresponds to a single occurrence of the event.
Define the event time-coverage of the re-occurring event $e$ to be:\\
\begin{equation*}
    TC(e) = \cup_i [t_i, t_i + l],\;\;\; i \in \{1, 2, \cdots\}
\end{equation*}
Then we define the basis function:
\begin{equation*}
    s_k(t;\;e) =
        \begin{cases}
            0 & t \notin  TC(e) \\
            \sin(2\pi(t - t_i)/l) & t \in TC(e)
        \end{cases}
\end{equation*}
Similarly, we define $c_k(t; e)$ as the cosine counterpart.\\
In order to model the effects of event $e$, we add the basis functions
\begin{equation*}
    \{s_k(t; e), c_k(t; e), k=1,\cdots,k_e\},
\end{equation*}
where $k_e$ is the appropriate Fourier Series order, to the set of basis functions. \\
Note that for a holiday $e$, the time coverage can be defined to include an
expanded window surrounding that holiday instead of solely the holiday.

Moreover, when $\{K=k_e\}$ (same Fourier order used for seasonality as for the holiday),
the basis functions can be expressed as interaction terms e.g.
\begin{eqnarray*}
s_k(t; e) =  s_k(t) * \mathbbm{1}_{TC(e)}, \\
\end{eqnarray*}
where $\mathbbm{1}_{TC(e)}$ is the indicator function.

\subsubsection*{Remaining Temporal Dependence}
After taking into account trends, seasonality,
events, change-points (discussed in the next section),
and other important features, often the residuals still
show a temporal dependence
(albeit often orders of magnitudes smaller than the original series).  
The remaining temporal correlation can be exploited
to improve the forecast especially for short-term forecast horizons.
We allow for an auto-regressive structure in the model e.g.\;by including lags in
the model: $Y(t-1), \cdots, Y(t-r)$ for some appropriate $r$.

While auto-regression can account for the remaining correlation in the series,
for many applications a large $r$
might be needed to capture long-term dependence in the chain.
To remedy this issue, \cite{hosseini-bin-pcpn}  suggested a 
technique to develop parsimonious models by aggregating the lags.
As an example, for a daily series, consider
the averaged lag series 
\[AVG(Y(t); 1,2,\cdots,7) = \Sigma_{i=1}^7 Y(t-i) / 7.\]
This covariate then represent the average value for the value of the series over the past week.
As another example, consider 
\[AVG(Y(t); 7,7\times2,7\times3) = \Sigma_{i\in (1, 2, 3)} Y(t-7i) / 3.\]
This covariate represent the average value of the series on the same day of week in the past 3 weeks.
Similar series can be defined for other frequencies as well.     

\subsubsection*{Lagged regressors and regressors}
Suppose $p$ other time series are provided with the same frequency as the
target time series to forecast:
$\bX(t) = X_1(t), \cdots, X_p(t)$.
If such other metrics are available while future values are unknown,
one can still use their lags in the models, similar to 
previous sections. 

In the case that the future values of the these variables are known,
or can be forecasted reliably based on other models, we can directly
use $X_1(t), \cdots, X_p(t)$ as regressors.
This also has applications in scenario-based forecasting where some underlying variables
are to behave differently into the future.

\subsubsection*{Accommodating complex patterns}
One of the advantages of our model compared
to many commonly used models in practice is the ability
to easily accommodate complex patterns.
This is done by specifying interactions (between features).
To mitigate the risk of the model
becoming degenerate or over-fitting,
regularization can be used in the machine learning
fitting algorithm for the conditional model (e.g.\;Ridge).
In fact, regularization also helps in minimizing the risk of divergence
of the simulations of future series for the model which are 
discussed in \cite{hosseini-bk-2020}.

Here we provide some examples of how complex patterns can be accommodated with interactions.
\begin{itemize}
    \item Growth is different on weekdays and weekends.
    This is possible and unobserved in many series of interest.
    For example it could be the case that the usage of a particular app surges much more
    rapidly on weekends as compared to weekdays.
    Suppose the time frequency is daily
    and the model includes the categorical term DOW to model the weekly patterns.
    To accommodate this pattern we can interact the growth term e.g.\;$f(t) = t$ with
    DOW:
    \[f(t) * DOW\]
    where $*$ denotes the interaction as commonly used
    in standard softwares such as R (https://www.R-project.org/)
    or Python's patsy package (https://patsy.readthedocs.io/).
    If a Fourier series is used to model weekly patterns,
    then one can interact the Fourier series terms or a subset of it with the 
    growth function.
    \item Different months have different patterns for day of week. 
    It might be possible that during different times of the year,
    the weekly patterns differ.
    Then one can use this interaction:
    \[f(t) * month\]
    where $month$ is a categorical variable denoting the month of the year.
    Similar interactions can be considered if Fourier series are used to model the annual trends.
\end{itemize}

\subsection{The volatility component}
\label{subsect:volatility}
As discussed in the introduction,
we elect to separately fit the conditional mean model and the volatility / error model.
Integrated models of course can be considered where the model includes all components
(e.g.\;\cite{hosseini-takemura-2015} or \cite{book-west-1997}).

In theory, the advantage of an integrated model is accounting for all uncertainty in one model.
However, we consider a two-component model
for the following reason:
(a) more flexibility in the mean components model in terms of features and algorithm;
(b) more flexibility in the
volatility model;
(c) considerable speed gain by avoiding computationally heavy Maximum Likelihood Estimation 
(e.g.\;\cite{hosseini-takemura-2015})
or Monte Carlo Methods 
(e.g.\;\cite{book-west-1997}).
These factors are very important in a production environment
where often fast and reliable forecasts are needed.
The increased reliability is due to (a) more stable / robust estimates of the 
model parameters (b) less chance of the forecasted values into the future 
to diverege to unreasonable values
(as reported by \cite{book-tong-1990} and \cite{hosseini-bk-2020}).

Here we discuss the details for the volatility model.
Suppose $Y(t)$ is the target series and $\Yhat(t)$ is the forecasted series.
Then define the residual series as follows:
\[R(t) = Y(t) - \Yhat(t)\]
Assume the volatility depends on given categorical features
$F_1, \cdots, F_p$ which are also known into the future.
As an example, these features could be the raw time features defined previously such as
``Day of week (DOW)'' or ``Is Holiday''
which determines if time $t$ is a holiday or not.
Then given any combination of features $F_1, \cdots, F_p$,
we consider the empirical distribution
$(R | F_1, \cdots, F_p)$ and fit a parametric or non-parametric distribution to the combination
as long as the sample size for that
combination, denoted by $n(F_1, \cdots, F_p)$ is sufficiently large
e.g.\; $n(F_1, \cdots, F_p) > N,\; N=20$.
Note that one can find an appropriate $N$ using data e.g.\;during cross-validation steps
by monitoring the distribution of the residuals.
Then from this distribution, we estimate the quantiles:
$Q(F_1, \cdots, F_p)$ to form the 95\% prediction interval:
\[\Yhat(t) + Q(F_1, \cdots, F_p)(0.025), \Yhat(t) + Q(F_1, \cdots, F_p)(0.0975),\]
and similarly for other prediction intervals. 
One choice for a parametric distribution is the Gaussian distribution
$\mathcal{N}(0, \sigma^2(F_1, \cdots, F_p))$
which can be appropriate for some uses cases.
The reason, that we have forced the mean to be zero is, we do 
not want to volatility model to modify the forecasted value which
is the result of often much more complex mean model with many more features.
Regarding normality assumption, note that while the original series can be heavily skewed,
it could be the case that the residual series 
is close to normal, especially after conditioning on $F_1, \cdots, F_p$.
For use cases where the normality of the conditional residuals
is not reasonable,
we can use non-parametric estimates of the quantiles using various approaches e.g.\;by simply
calculating sample quantiles for sufficiently large samples.

We also need to address the case where the sample size for a given combination is smaller than $N$.
In this case, we need to fall back to another reasonable way to come up with prediction interval.
To determine the fall back values: we calculate the Interquartile Range $IQR(c) = (Q(0.75) - Q(0.25))$ 
for each combination $c$; then order these IQR increasingly and for a large $p \in [0, 1]$, e.g.\;$p=0.9$, 
we pick $c_0$ which attains that IQR. Then we use the quantiles of $c_0$ as the fall back value.
The idea behind this approach is to fall back to some quantiles which are coming from more variable combinations.
Note that in practice, the combinations should be chosen in a way that no fall-back is necessary and this mechanism is 
only there to cover rare cases when a combination gets assigned smaller sample sizes. For future work, one can also consider
parametric models for volatility in residuals to be able to accommodate a large number of features in volatility. 
The concrete steps for fitting the volatility model are described in Algorithm \ref{alg:volatility}.

\vspace{1cm}
\begin{algorithm}[H]
    \label{alg:volatility}
    \SetAlgoLined
    \KwData{Apply the forecast model to generate predictions}
    \SetKw{Pp}{Calculate residuals}
    \Pp{\\}
        {
        \Indp
            
        }
    \SetKw{Rg}{Split data based on feature values}
    \Rg{\\}
        {
        \Indp

        }
    \SetKw{Pop}{Estimate distribution for large combinations}
    \Pop{\\}
        {
        \Indp
            Consider combinations $C_{large}$ with sample size $\geq N$\;
            \For{combination $c$ in $C_{large}$}{
                Calculate $IQR(c)$\\
                Calculate desired lower and upper bounds: $q_1(c),q_2(c)$
            }
        }
    \SetKw{Pop}{Estimate fall-back volatility for small combinations}
        \Pop{\\}
        {
            \Indp
            Order the combinations $C_{large}$ with respect to $IQR(c)$.\\
            Find the combination, $c_0$, which is percentile of $p$ in $C_{large}$.\\
            Set fall back quantiles to $q_1(c_0), q_2(c_0)$.\\
            Consider combinations $C_{small}$ with sample size $< N$\;
            \For{combination $c$ in $C_{small}$}{
                Set lower and upper bounds to the fall back quantiles
            }
        }
    \KwResult{Return $(q_1(c), q_2(c))$ for all $c$; and fall back quantiles $q_1(c_0), q_2(c_0)$}
    \caption{Volatility algorithm}
\end{algorithm}

    \section{Changepoint Detection}
\label{sect:change-points}






Changepoints play an important role in forecasting problems.
By a changepoint,  we refer to a time point in a time series,
with the pattern in the data segment after the time point exhibiting a change
from the data segment prior to it.
Capturing the changepoints can help the model adapt to the
most recent data patterns and learn the right behavior in forecasts.
This section discusses a changepoint detection algorithm
that focuses on trend changepoints.
The algorithm is based on the adaptive lasso \citep{zou2006adaptive}
to select significant changepoints from a large number of potential changepoints.
While fully automatic detection is possible, tuning parameters are provided to allow flexibility.

\subsection{Trend Changepoint Detection}
\label{subsect:trend-change-points}

The trend of a time series describes
the long-term growth that ignores seasonal effects or short-term fluctuations.
As discussed in Subsection \ref{subsubsect:growth},
in general the trend can be denoted as a function of time, i.e.,
$$\text{long-term trend}=g(t)$$
for a continuous function $g$.
This growth function can be approximated with basis functions such as
linear, quadratic, cubic and logistic.
Here, we focus on the case with linear basis functions
as it suffices to be a good starting point for most applications we have encountered.
In fact, according to the Stone-Weierstrass Theorem
(Chapter 7 of \cite{book-rudin-1976}), any continuous
function can be approximated by a piece-wise linear functions.
(Note however, this does not imply that there are
no merits in considering other basis functions in other applications,
as other basis functions could
potentially capture some complex trends more parsimoniously,
but we do not discuss that here further for brevity.)

Given a list of changepoints $t_1,\cdots,t_k$,
we can approximate the continuous function $f$ with piece-wise linear function
$$g(t)=a_0t + \sum_{i=1}^ka_i\indicator_{\{t>t_i\}}(t-t_i).$$

The approximation error gets smaller when we have more changepoints,
while the risk of over-fitting also gets higher, thus making the forecast less reliable.
Therefore, the method discussed here intends to only select significant trend changepoints.
The proposed method here is a regularized regression based algorithm with
a few filters being applied for practical considerations,
and is able to identify significant trend changepoints.
The steps of this procedure are given in algorithm \ref{alg:trend}.

First, we apply an optional aggregation.
For example, aggregating daily data into weekly data.
The main purpose of applying this aggregation is to eliminate short-term fluctuations.
For example, a short holiday effect on daily data should not be picked up
as long-term trend changes. Note that this aggregation process
may not be necessary if data frequency is already sufficiently coarse e.g.\;for weekly data.

The next step is to place a fine regular grid of potential changepoints uniformly over
the span of time of the series.
For example for daily data with weekly aggregation,
placing potential changepoints every week or every two weeks can be considered.
However, it also would introduce false or pseudo changepoints.
Therefore, a balanced approach is needed, which we will consider shortly.
Large number of potential changepoints,
also guarantees the existence of
a potential changepoint, sufficiently close to
any ``true'' trend changepoint.

We allow for some other restrictions in the procedure to find the appropriate change points.
For example, we allow for the user to specify a period at the end of the series
where no change point is allowed.
The reasoning is that the position of last changepoint and its slope afterwards has a significant
impact on the forecasted values.
This way, we allow for expert knowledge of the nature of the series of interest to be
incorporated into the model.

The core step applies the Adaptive Lasso \citep{zou2006adaptive}
to the regression problem with the aggregated time series
as the response and the changepoints and yearly seasonality as regressors, i.e.,
\begin{equation}
\label{eq:adalasso}
y_{t, agg} = a_0t + \sum_{i=1}^ka_i\indicator_{\{t>t_i\}}(t-t_i) + \sum_{j=1}^K\left(\beta_{ci}c_j+\beta_{si}s_j\right),
\end{equation}
where $c_j$ and $s_j$ are the sin and cosine functions corresponding to the
yearly Fourier series,
as discussed in Subsection \ref{subsubsect:seasonality},
$$ s_k = \sin (k \omega_y y(t)),\;c_k = \cos (k \omega_y y(t)), \;\omega_y = 2 \pi, k=1,\cdots,K_1,$$
where $y(t)$ denotes TOY (time of year) defined in Subsection \ref{subsect:mean-component}.
The adaptive lasso adds weights to the $L_1$ norm penalization of the lasso \citep{tibshirani1996regression},
where the weights are chosen based on various rules, 
for example, the reciprocal of some initial estimations of the coefficients.
The reason that we use the adaptive lasso over the lasso
is that the adaptive lasso can gain the desired sparsity level
without over-shrinking the significant coefficients,
with properly chosen weights,
as discussed in \cite{zou2006adaptive}.
If the time series has a long history,
yearly seasonality change may be falsely captured as trend change,
so it's better to refit yearly seasonality after some period, e.g., every 2 years.
In the regression formulation above,
this can be handled by introducing extra regressors that account for yearly seasonality change, e.g.,
$$s_{k1} = \indicator_{\{t>t_1\}}\sin (k \omega_y y(t)),\;c_k = \indicator_{\{t>t_1\}}\cos (k \omega_y y(t)), \;\omega_y = 2 \pi, k=1,\cdots,K_1,$$
for some change period $t_1$.
Introducing theses terms will fit different yearly seasonality before and after $t_1$.

Note that we should only penalize the changepoint parameters i.e.,
\begin{align}
\label{eq:partialreg}
&\hat{a}_0, \cdots, \hat{a}_k, \hat{\beta}_{c1}, \hat{\beta}_{s1}, \cdots, \hat{\beta}_{cK}, \hat{\beta}_{sK}\nonumber\\
=&\argmin\sum_{m=1}^n\left[y_{t,agg}-\left(a_0t + \sum_{i=1}^ka_i\indicator_{\{t>t_i\}}(t-t_i) + \sum_{j=1}^K\left(\beta_{ci}c_j+\beta_{si}s_j\right)\right)\right]^2+\lambda\sum_{i=1}^kw_i|a_i|,
\end{align}
for some weights $w_i$, $i=1,...,k$.
This partial penalized regression can be solved
with a fast coordinate descent algorithm \citep{tseng2001convergence} .
An alternative is to use projection to split the optimization problem into two steps.
The first step fixes the $L_1$-norm penalized coefficients and estimate
the other coefficients as a ridge regression problem.
The second step uses the estimated coefficients from the first step and reduce
the problem into a Lasso regression problem.
Both methods requires some math and programming.
A derivation of the algorithm is given in appendix \ref{subsect:mix-regularization}.
It is worth noting that
we are not losing too much by directly optimizing (\ref{eq:adalasso})
over (\ref{eq:partialreg}).
The only difference is that penalizing the $\boldx_0$ term
brings in a pseudo-changepoint at the very beginning of the time series to
``pick up" the baseline trend.
This can be avoided by not allowing potential changepoints
to be placed near the very beginning of the time series.
Existing libraries such as \textit{sklearn} \citep{pedregosa2011scikit} can be utilized
to solve the problem efficiently.


\begin{algorithm}
\label{alg:trend}
\SetAlgoLined
\KwData{Time series data.}
\SetKw{Pp}{Pre-processing:}
\Pp{\\}
{
\Indp
    Aggregation (optional)\;
    Put a large number of potential changepoints uniformly\;
    Eliminate the potential changepoints in forbidden region (e.g.\;the end)\;
}
\SetKw{Rg}{Regularization:}
\Rg{\\}
{
\Indp
    Create trend features (including changepoints) and yearly seasonality features\;
    Fit a regression model with the features (initial estimation)\;
    Compute feature weights\;
    Fit regression model with adaptive l1 norm penalty on the changepoint features\;
}
\SetKw{Pop}{Post-pressing:}
\Pop{\\}
{
\Indp
    Group selected changepoints with distance less than a threshold\;
    \For{group in groups}{
        \While{can add or remove changepoints}{
            Remove the changepoint with smaller change if two or more are too close\;
            Trace back to see if any changepoint can be added back\;
        }
    }
}
\KwResult{The detected trend changepoints.}
\caption{The automatic trend changepoint detection algorithm.}
\end{algorithm}


After the regularization,
we have selected significant trend changepoints.
However, there can be detected changepoints that are too close to each other.
These types of changes usually account for consecutive changes or gradual changes.
As for pure trend changepoint detection, the result does not need further handling.
If one wants to use these changepoints as input for a forecast model and prefers parsimony,
a rule-based post-filtering method can be applied to remove redundant trend changepoints.

Here we provide more details about filtering redundant change points. First, we group changepoints
that are close to at least one another with respect to some threshold.
For each group, we start with the first changepoint and look at the second one.
The one with smaller changes are dropped, and we go the the third one, and so on.
If we dropped the second changepoint,
and the distance between the first and the third changepoints is greater than the threshold,
we won't need to compare those two changepoints,
and will continue to compare the third and the fourth changepoints.
In a special case when $c_1<c_2<c_3$,
where $c_i$ is the magnitude of the $i^{th}$ changepoint,
after dropping the first and the second changepoints,
the first changepoint can be added back,
if the distance between the first and the third changepoints is large enough.
Each time we remove a changepoint,
we check back to see if any deleted changepoints can be added back.
This rule-based filtering method not only enforces minimal distance
to the detected trend changepoints,
but also retains as many changepoints as possible.

Through the whole algorithm, the following components can be customized to fit specific use cases
\begin{itemize}
    \item The aggregation frequency.
    \item The distance between potential changepoints or how many potential changepoints.
    \item The time period(s) with no changepoints.
    \item The yearly seasonality Fourier series order.
    \item How often yearly seasonality changes.
    \item The initial estimation method for adaptive Lasso.
    \item The regularization strength.
    \item The minimum distance between detected changepoints.
    \item Any customized trend changepoints to be added to detected changepoints.
    \item The minimum distance between a detected changepoint and a customized changepoint.
\end{itemize}

The above algorithm can stand alone as a trend changepoint detection algorithm.
It runs fast and gives significant trend changepoints to help capture
long-term events, product feature launches, changed in underlying dynamics, etc.
On the other hand,
the output can also be used to specify the trend changepoints as in Section \ref{sect:model}.
The advantage of detecting trend changepoints independently from performing it
while fitting the model is to allow more flexible algorithms in the fitting phase,
because it separates estimation for more interpretability,
and makes it easier to tune sub-modules.

\subsection{Seasonality Changepoint Detection}
\label{subsect:seasonality-change-points}

In some time series, seasonality effect can change over time similarly to trend.
Product features and news events can increase or decrease the volatility of the time series.
Gradual changes in volatility can be modeled with interactions,
however, seasonality changepoints are a more flexible and automatic way to capture this effect.

Seasonality effect is modeled with Fourier series terms in the model,
and this makes it easy to include seasonality changepoints.
We use truncated seasonality features to capture seasonality changes.
For every Fourier series term $s_k$ and $c_k$,
the truncated terms $s_k\indicator_{\{t>t_{scp}\}}$ and $c_k\indicator_{\{t>t_{scp}\}}$
models the change after a seasonality changepoint $t_{scp}$.
A regularized regression can be used to choose significant seasonality changepoints
from a large number of potential seasonality changepoints as we do in
automatic trend changepoint detection.

The main difference between trend changepoints and seasonality changepoints lies on two parts.
The first part is that the seasonality has multiple components,
for example,
yearly seasonality,
quarterly seasonality,
monthly seasonality,
weekly seasonality,
daily seasonality, etc.
The second part is that for each component,
there are multiple Fourier series bases that account for the same seasonality changepoint.

In our model, a group lasso type of regularization \citep{yuan2006model} is
used on the the Fourier bases for each potential changepoint of each component.
This method drops all Fourier bases of one changepoint in one component entirely,
however, the resulted seasonality is not smooth because of the entire changes.
Using $L_1$-norm over all terms can still give the desired sparsity level
and also gives smoother seasonality changes.
Therefore, the same adaptive method discussed in trend changepoints
also works for seasonality changepoint detection in practice.

    \section{Example application to bike-sharing data}
\label{sect:use_case}

This section demonstrates how various model components of our proposed model work
through an example. We consider a dataset with hourly counts of rented bikes in a bike-sharing system.

The bike-sharing data include the hourly counts of rented bikes in
Washington DC during 2011 to 2019 and were obtained from
\textit{www.capitalbikeshare.com}.
We have also joined this data with daily weather data
from a nearby station (BWI Marshall Airport).
The weather data was downloaded from
``Global Historical Climatology Network'' (\textit{https://www.ncdc.noaa.gov/}) and
contains the daily minimum and maximum temperature and precipitation total.

\begin{figure}[H]
    \centering
    \includegraphics[width=\textwidth]{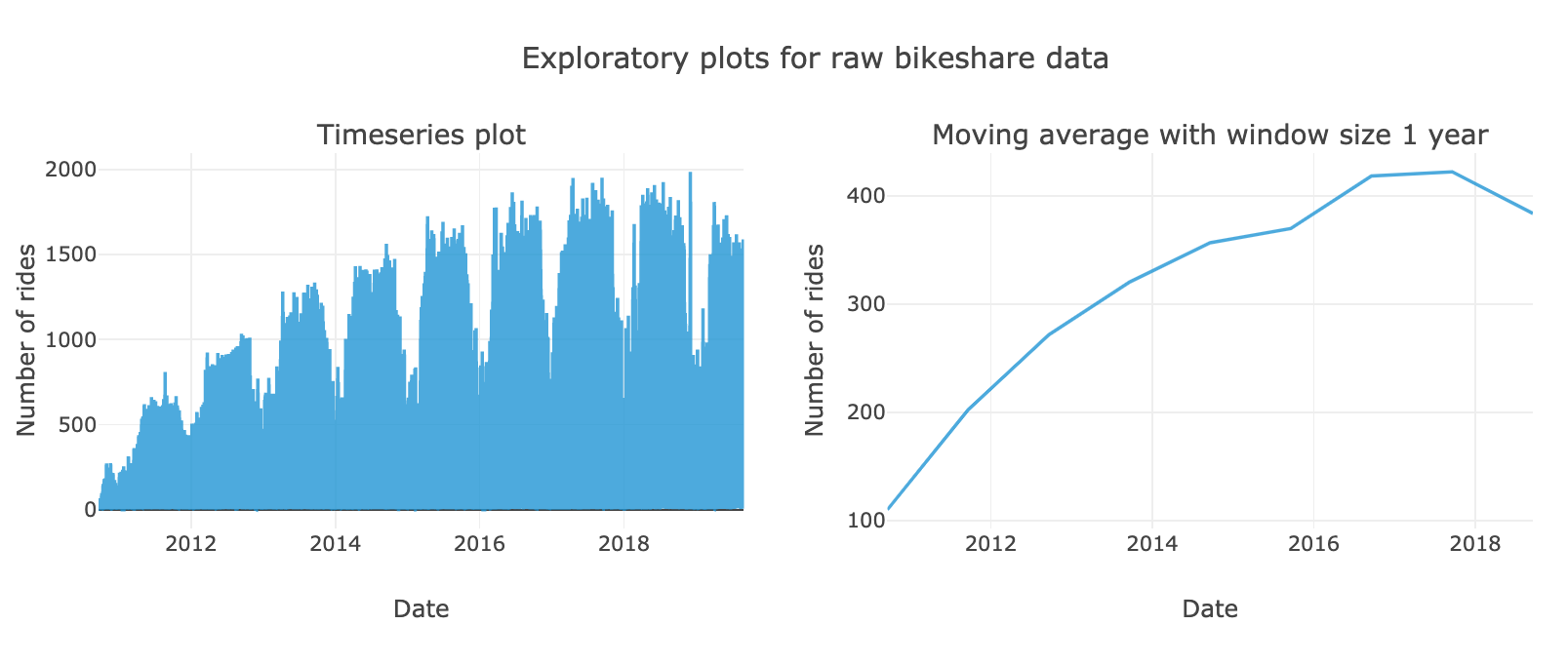}
	\caption{The raw time series (left) and
        the trend by a moving average of 365 days
	    (right).}
    \label{fig:bikeshare_rawplot}
\end{figure}

Figure \ref{fig:bikeshare_rawplot} shows the time series.

To build a Silverkite model, we will go over the components
and try to find out what are the
key components to be used in the model.
First, the Right Panel in Figure \ref{fig:bikeshare_rawplot}
shows the rolling mean of
the raw time series with a rolling window size of $24 \times 365$ (1 year).
We observe that the trend is roughly increasing before late 2017 and starts
to decrease after that.
There are some slight growth rate changes but overall the trend can be modeled
with a linear trend with changepoints.

\begin{figure}[H]
	\centering
	\includegraphics[width=1.0\textwidth]{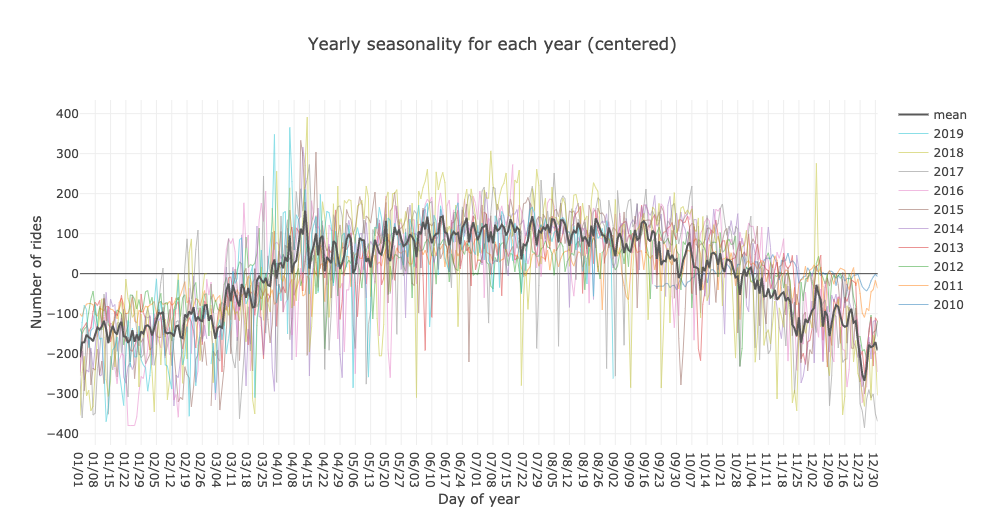}
	\caption{Yearly seasonality of the bike-sharing data.}
	\label{fig:bikeshare_yearly}
\end{figure}

\begin{figure}[H]
	\centering
	\includegraphics[width=1.0\textwidth]{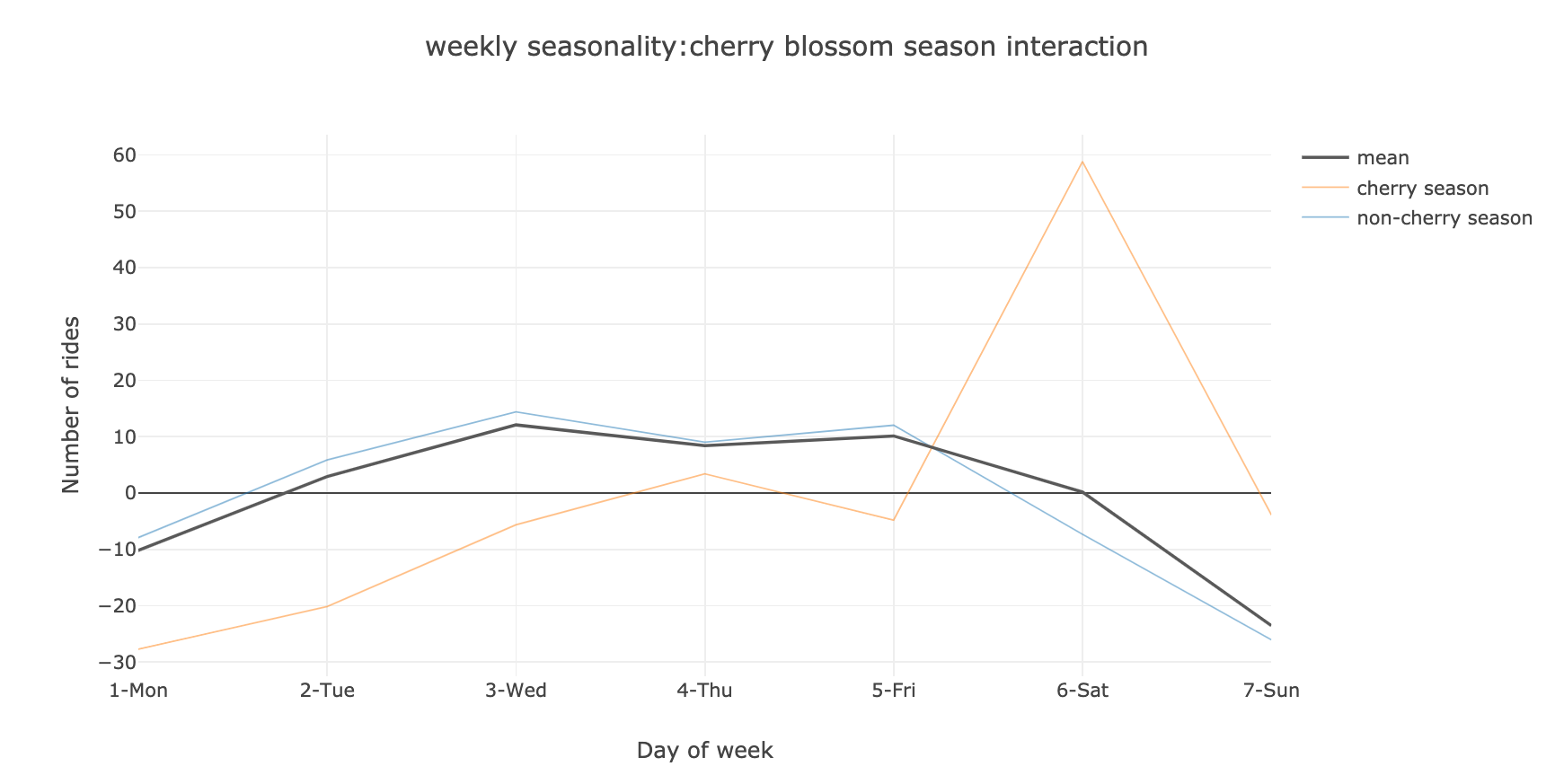}
	\caption{Different weekly seasonality for cherry season (Mar 20-Apr 30) and non-cherry season.}
	\label{fig:bikeshare_weekly_cherry}
\end{figure}

The seasonality components include different length of periods.
We mainly focus on yearly seasonality, weekly seasonality and daily seasonality,
because the other periods do not have apparent patterns or convincing seasonal reasons.
Figure \ref{fig:bikeshare_yearly}
shows the yearly seasonality which indicates that there are more rides during warm months.

From the "mean" line in Figure \ref{fig:bikeshare_weekly_cherry}, the weekly seasonality
is not very clear at first.
However, upon inspecting the weekly seasonality broken down by month,
we observe that the weekly seasonality in April differs significantly from other months.
The reason is that Washington DC has one of its most significant events
around April -- The Cherry Blossom Festival.

Figure \ref{fig:bikeshare_weekly_cherry} shows the contrast in
weekly seasonality pattern
between the Cherry Blossom Season (approximately March 20 -- April 30)
and the rest of the year.
This observation can be modeled using an interaction
between weekly seasonality and an indicator variable determining
if the observation is in Cherry Blossom Season or not.

\begin{figure}[H]
	\centering
	\includegraphics[width=1.0\textwidth]{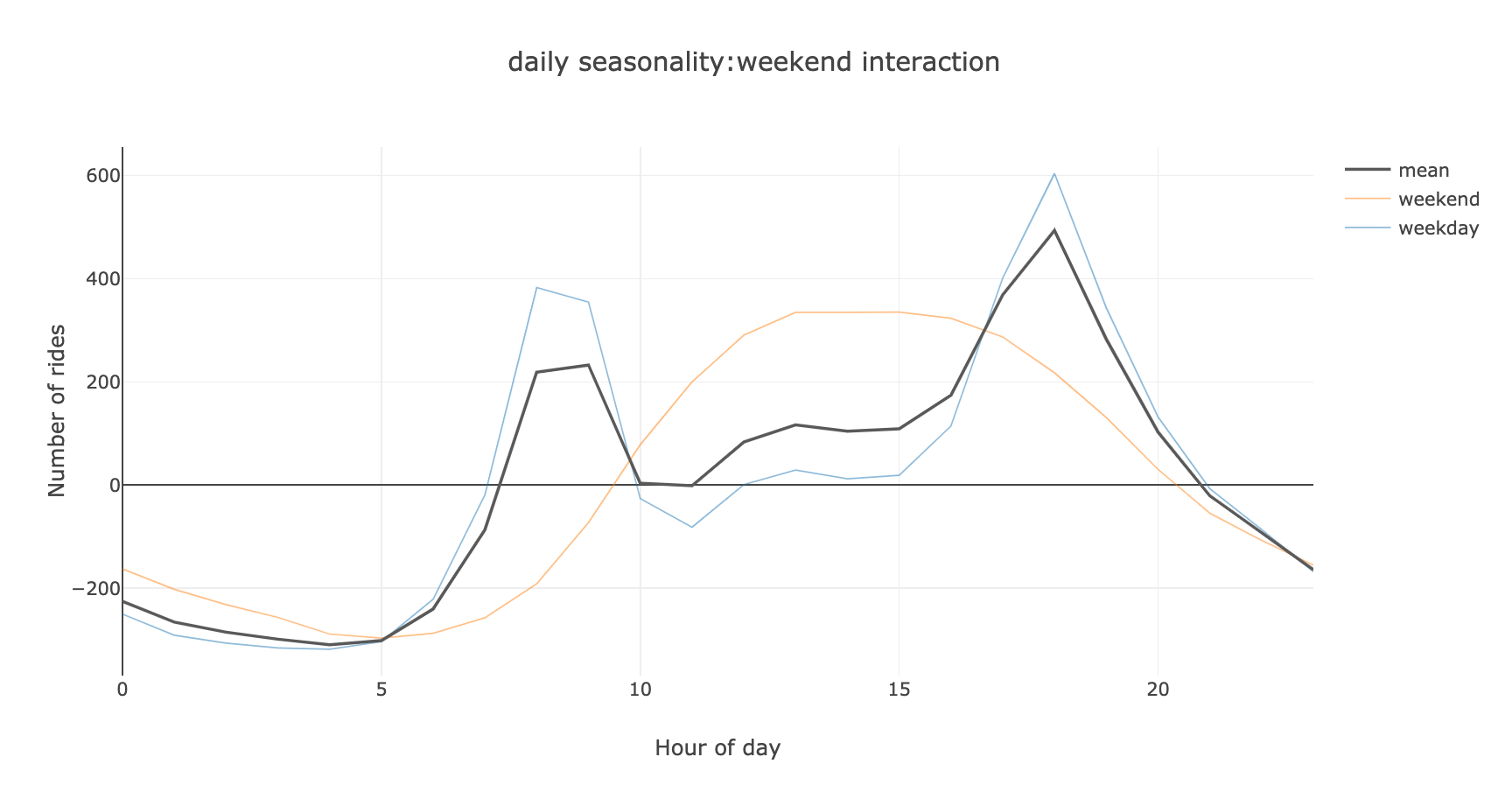}
	\caption{Different daily seasonality for week-days and weekend.}
	\label{fig:bikeshare_daily_weekend}
\end{figure}

The daily seasonality pattens are very significant but differ
on weekdays versus weekends. This is typical for time series that
involve human activity.
The daily seasonality effect interacting with weekend is shown in
Figure \ref{fig:bikeshare_daily_weekend}.
The figure suggests to differentiate between daily patterns
of weekdays and weekends.
This can be done in various ways, for example, by allowing for
an interaction between seasonal daily patterns
and a weekend indicator.
Alternatively we could consider a sufficiently
large number of Fourier series terms for
a weekly pattern (or a combination of both).

The holiday effect is hard to see from component plots,
but from Figure \ref{fig:bikeshare_yearly},
we can see there are some small dips that are due to holidays.


In summary, the models will include the following components
\begin{itemize}
    \item CT (continuous time to model the long-term trends)
    \item changepoints
    \item yearly seasonality
    \item weekly seasonality and its interaction with Cherry Blossom Season
    \item daily seasonality : weekend interaction
    \item holidays
    \item weather regressors
    \item autoregression
\end{itemize}

First we consider a forecast horizon of $24 \times 14$ (2 weeks).
Because the number of rides is always non-negative,
we clip negative values at zero.
A linear growth is used as long-term trend function.
For trend changepoints,
we create a grid of potential changepoints every 15 days but skip the last 30 days
(to avoid detecting artificial change points toward the end of the series).
A yearly seasonality of order 15 is used in the changepoint detection algorithm.
For the Lasso problem,
it's easy to verify by KKT conditions that
$\|\boldX^T\boldy\|_{\infty}/n$ is the minimal
tuning parameter that corresponds to no non-zero coefficients.
The regularization parameter, we use in detecting changepoints is $10^{-3}$ of that value,
which provides moderate trend changepoints.
An aggregation of 3 days is used for the changepoint detection aggregation process
(Algorithm \ref{alg:trend}).


For seasonality, we use order 15, 3 and 12 for yearly,
weekly and daily seasonality, respectively.
Because the volatility changes over time as well,
we also introduce seasonality changepoints,
which allows seasonality components to refit after changes, however,
the seasonality changepoints is not allowed within the last 365 days of data,
to avoid a potentially poor fit of yearly seasonality.

A list of common holidays together with their plus/minus 1 day were created as separate indicators.
We also include the two interaction groups we discussed above:
weekly seasonality and Cherry-Blossom Season indicator;
daily seasonality and weekend indicator.

\begin{figure}[H]
	\centering
	\includegraphics[width=1.0\textwidth]{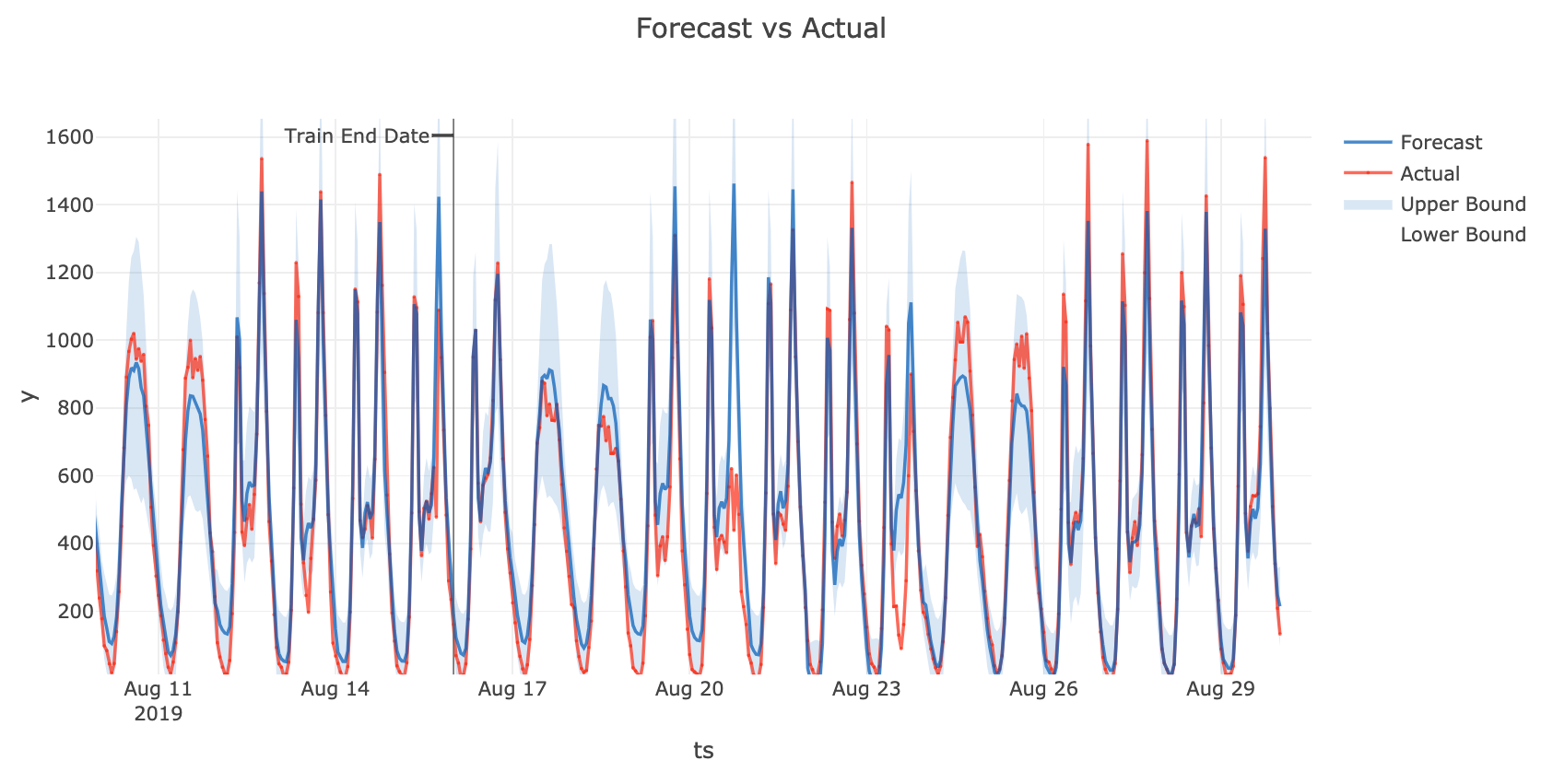}
	\caption{Forecast versus actual for Silverkite on bike-sharing data
        during test period.}
	\label{fig:bikeshare_forecast}
\end{figure}

\begin{figure}[H]
	\centering
	\includegraphics[width=1.0\textwidth]{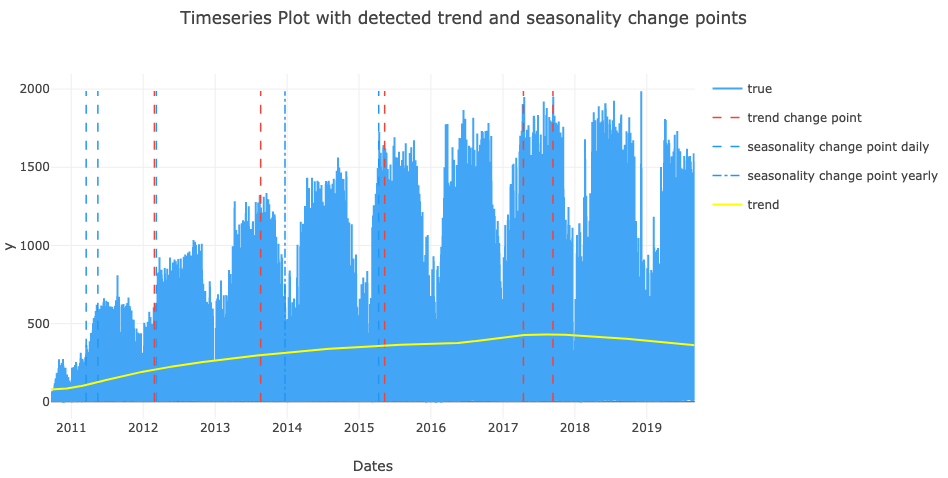}
	\caption{Detected trend and seasonality changepoints for bike-sharing data.}
	\label{fig:bikeshare_trend_change}
\end{figure}

\begin{figure}[H]
	\centering
	\includegraphics[width=1.0\textwidth]{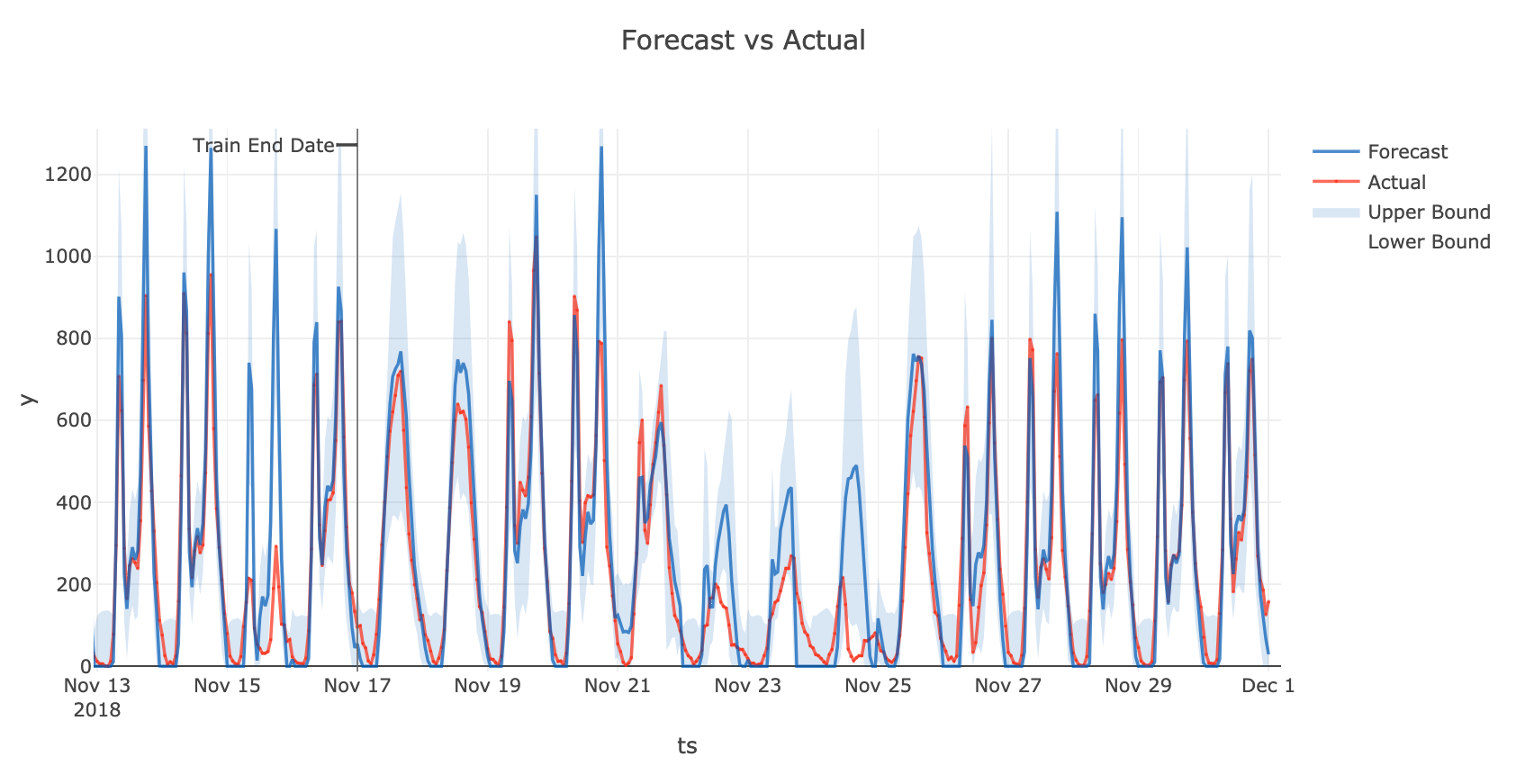}
	\caption{Forecast versus actual for Silverkite
        on bike-sharing data during Thanksgiving.}
	\label{fig:bikeshare_forecast_holiday}
\end{figure}

Figure \ref{fig:bikeshare_forecast} compares
the forecast with the actual data during the test period,
which is located at the end of the time series.
We can see that with our daily seasonality and weekend indicator interaction,
the model is able to capture the bimodal daily seasonality on weekdays
and unimodal daily seasonality on weekends.
Figure \ref{fig:bikeshare_trend_change}
shows the detected trend and seasonality changepoints and the estimated trend in the model.
It aligns with our expectations.
Figure \ref{fig:bikeshare_forecast_holiday}
shows the forecast versus the actual data around Thanksgiving 2018.
From the figure, we see the holiday effect is picked up by the model,
by observing that the number of rides is lower compared to regular
non-holiday number of rides on the same weekday.
Moreover, the interaction between whether
it is weekend and daily seasonality also captures the uni-modal shape
on these holidays (non-workdays).
A gap in prediction happens on Nov 15 in the same plot.
The reason is that Washington DC was hit with biggest November snowfall in 29 years,
and people wasn't ready for a snow riding commuting yet.
We weren't explicitly capturing this effect so there is the gap,
however, any knowledge-based observations can be modeled with extra regressors.

Autoregressive components are useful in picking up remaining trends which are not
explained by seasonality, growth, events and change points.
The effect of autoregression cannot be easily observed from figures.
A cross-validation study is used to demonstrate its usefulness.
The cross-validation includes 20 folds.
The validation folds are taken from the end of the data set,
with a 2-week window between each fold.
The forecast horizon is set to 24 hours for short-term use case.

From the cross-validation study, we have an RMSE of 102.3
when including autoregressive components,
compared to 123.0 when not including autoregressive components.
Let $Y(t)$ be a time point in the forecast phase,
the autoregresive terms used include
$Y(t-24)$, $Y(t-25)$, $Y(t-26)$
as autoregressive terms.
They also include these three aggregated terms:
\[AVG(Y(t); 7 \times 24, 14 \times 24, 21 \times 24),\]
which is taking an average of observed values in the same hour of last three weeks.
\[AVG(Y(t); 24, 25, \cdots, 191),\]
which is the average of the last 7 days;
and
\[AVG(Y(t); 192, 193, \cdots, 359)\]
which is the average of the week prior to last.
These are the default choices (for hourly data with forecast horizon of 24)
in our model and are not optimized
for this use case in particular.
However, the user can use grid search to optimize these choices further.


Note that the minimum lag used in the model is 24 which is
the forecast horizon. This will assure that, at prediction time,
all the lags used in the model to forecast are observed and not simulated.
This sometimes can help with accuracy, and is also beneficial in terms of
speed as we do not need to perform simulations at the prediction phase
(to fill in the lags needed into the future).
It is worth mentioning however, that the optimality of such choices
depend on underlying use case and user should experiment with various
models.

    \section{Assessment and Benchmarking}
\label{sect:assessment}
This section presents details on appropriate methods to assess the performance of
forecasting algorithms in terms of accuracy.

To estimate how accurately a forecasting model (e.g.\; Silverkite) performs in practice,
we use cross-validation (CV).
Cross-validation is a technique for assessing how
the results of a predictive model generalizes to
new data. In the time series context the new data refers to values in the future. Below
we describe an assessment method which is appropriate for forecasting applications.

We benchmark the prediction accuracy of Silverkite against
that of popular state of the art algorithms such as
Auto-Arima and (Facebook) Prophet.
The prediction accuracy is measured by
Mean Average Percentage Error (MAPE) which is defined as
\begin{equation}
    \text{MAPE}(Y) = \sum_{t=1}^T \Bigg \lvert \frac{Y(t) - \hat{Y}(t)}{Y(t)} \Bigg \rvert
\end{equation}
MAPE is more popular in applications as compared to RMSE (Root Mean Square Error), as it
provides a relative (scale-free) measure of error compared to the observation.

We use a rolling window CV for our benchmarking,
which closely resembles the well known $K$-fold CV method.
In $K$-fold CV, the original data is
randomly partitioned into $K$ equal sized subsamples.
A single subsample is held out as the validation data,
and the model is trained on the remaining $(K-1)$ subsamples (\cite{book-hastie-2009}).
The trained model is used to predict on the held-out validation set.
This process is repeated $K$ times so that each of the $K$ subsamples
is used exactly once as the validation set.
Average testing error across all the $K$ iterations provides
an unbiased estimate of the true testing error of the machine learning (ML) model on the data.

Due to the temporal dependency in time-series data the standard $K$-fold CV is not
appropriate.
Choosing a hold-out set randomly has two fundamental issues in time series context:
\begin{enumerate}
    \item Future data is utilized to predict the past.
    \item Some time series models can not be trained realistically with a random sample,
    e.g. the autoregressive models due to missing lags.
\end{enumerate}

\begin{figure}[H]
    \centering
    \includegraphics[width=\textwidth]{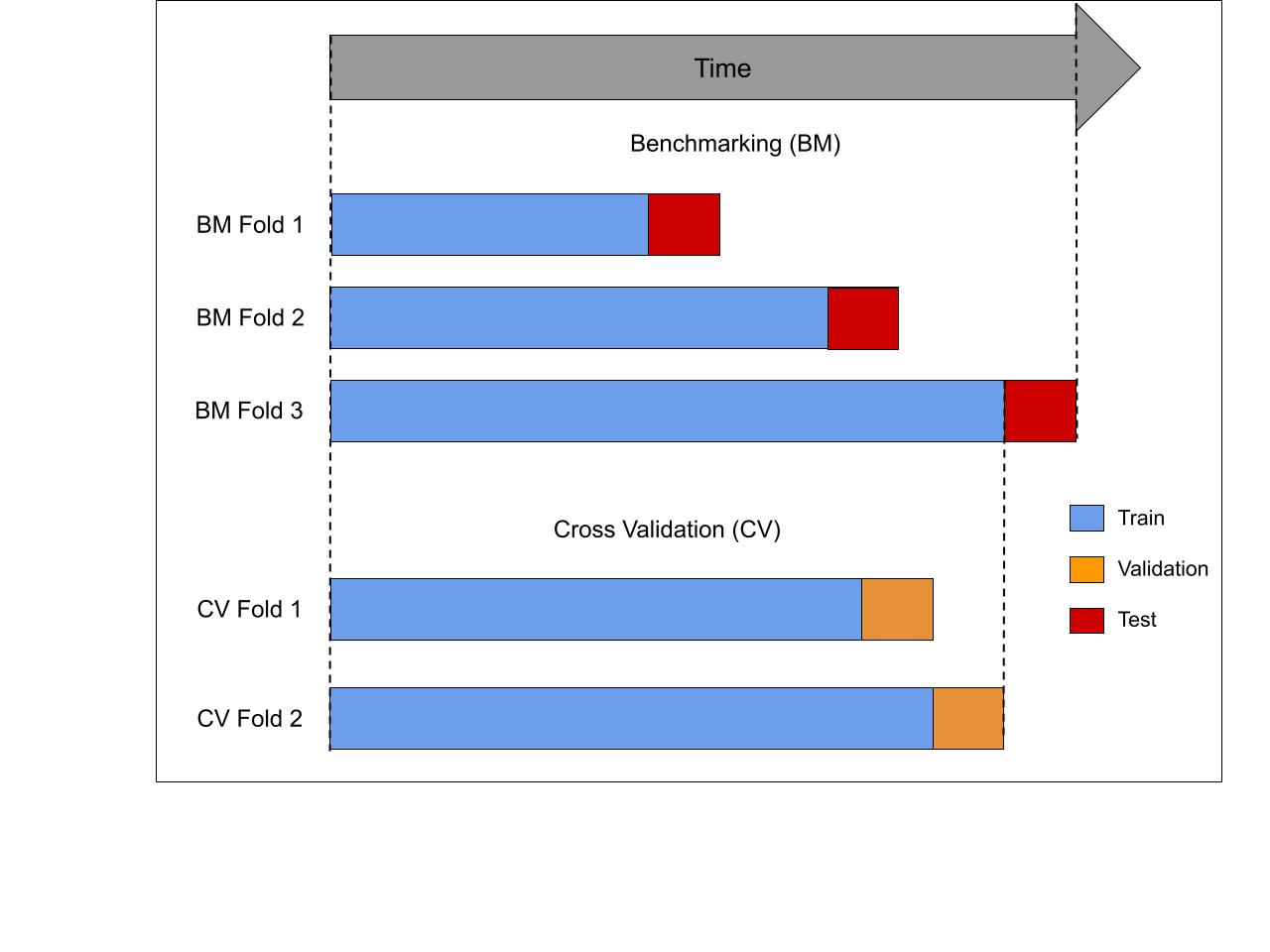}
	\caption{Fold structure for a Rolling Window Cross Validation.
    Expanding window configuration shown,
    where each training set has the same train start date.}
    \label{fig:benchmark_cv}
\end{figure}

Rolling window CV addresses this by creating a series of $K$ test sets.
For each test set, the observations prior to the test set are used as
a training set.
Within each training set, a series of CV folds are created, each
containing a validation set.
For time consideration, we choose to create a single CV fold for
each training set.
Number of data points in every test and validation set equals forecast
horizon (CV horizon in Table \ref{tab:benchmark-cv}).
Observations that occur prior to that of the validation set
are used to train the models for the corresponding CV fold (Figure \ref{fig:benchmark_cv}).
Thus, no future observations can be used in constructing the forecast,
either in validation or testing phase. The parameters minimizing
average error on the validation sets are chosen.
This model is then retrained on the training data for the corresponding becnh-mark (BM) fold.
The average error across all test sets provides a robust estimate of the model performance
with this forecast horizon.

\begin{table}[H]
    \tiny
    \centering
    \begin{tabular}{|c|c|c|c|c|c|}
    \hline
    \textbf{Frequency} & \textbf{\begin{tabular}[c]{@{}c@{}}Forecast \\ horizon\end{tabular}} & \textbf{\begin{tabular}[c]{@{}c@{}}CV\\ horizon\end{tabular}} & \textbf{\begin{tabular}[c]{@{}c@{}}CV minimum \\ train periods\end{tabular}} & \textbf{\begin{tabular}[c]{@{}c@{}}Periods \\ between splits\end{tabular}} & \textbf{\begin{tabular}[c]{@{}c@{}}Number \\ of splits\end{tabular}} \\
    \hline
    daily  & 1  & 1  & 365*2  & 25 & 16\\
    \hline
    daily    & 7   & 7    & 365*2    & 25  & 16   \\
    \hline
    \end{tabular}
    \caption{Details about the parameters used for cross-validation (CV) during benchmarking}
    \label{tab:benchmark-cv}
\end{table}

All the models are run on the $2$ different forecast horizons (1 day and 7 day) for daily data sets (Table \ref{tab:benchmark-cv}).
These horizons roughly represent short-term, average-term forecasts
for the corresponding frequency.
We plan to publish more bench marking results on more data sets, more frequencies (e.g.\; hourly, weekly)
and more time horizons in the future.

We require the datasets to have at least $2$ years worth of training data
so that the models can accurately estimate yearly seasonality patterns.
The number of periods between successive test sets and total number of
splits are chosen for each frequency to ensure the following:
\begin{enumerate}
    \item The predictive performance of the models are measured over an year
    to ensure that cumulatively the test sets represent real data across
    time properties e.g.\;seasonality, holidays etc.
    For daily data, $\text{periods between splits} (25)*\text{max splits}(16) = 400 > 365$,
    hence the models are tested over a year.
	\item The test sets are completely randomized in terms of time features.
    For daily data, setting ``periods between splits'' to any multiple of $7$
    results in the training and test set always ending on the same day of the week.
    This lack of randomization would have produced a biased estimate
    of the prediction performance.
    Similarly setting it to a multiple of $30$ has the same problem for day of month.
    A gap of $25$ days between test sets ensures that no such
    confounding factors are present.
	\item Minimize total computation time while maintaining the previous points.
    For daily data, setting ``periods between splits'' to $1$ and maximum number of splits to
    $365$ is a more thorough CV procedure.
    But it increases the total computation time $25$ fold and hence is avoided.
    We chose consistent benchmark settings suitable for all algorithms, including
    the slower ones.
\end{enumerate}

We have used out-of-box configuration for Auto-Arima and (Facebook) Prophet and Silverkite.
Silverkite uses a ridge regression to fit the model and contains
linear growth,
appropriate seasonality (e.g.\;quarterly, monthly and yearly seasonality for daily data),
automatic changepoint detection,
holiday effects, autoregression, and daily and weekly seasonality
interaction terms with trend and changepoints.

The benchmarking was run on three datasets for every forecast
horizon.
\begin{itemize}
    \item Peyton-Manning Dataset from fbprophet package (Facebook Prophet)
    \item Daily Australia Temperature Dataset, Temperature column
    \item Beijing PM2.5 Dataset
\end{itemize}

For consistency, no regressors were used for any dataset.
The entire process is executed on a system equipped
with 64 GB RAM and Intel Xeon Silver 4108 CPU @1.80 GHz.
The CPU has 8 cores, each with 2 threads.
The average
test MAPE and runtime across datasets are summarized in
Table \ref{tab:benchmark-summary}.
We can see that the default Silverkite MAPE is on par
with Auto-Arima in short-term forecasts, outperforms other algorithms in average-term forecasts.
However, Silverkite has a
clear speed advantage over Prophet. This makes prototyping quicker in Silverkite
and aids in building a customized more accurate model.

Note that the average test MAPE values are high due to values close to 0 in the Beijing PM2.5 dataset.
These are a few benchmarks on public datasets.
We plan to include more datasets, forecast horizons, and data frequencies that better match our industry applications and publish those results in the future.

\begin{table}[H]
    \tiny
    \centering
    \begin{tabular}{|c|c|c|c|c|c|c|l|}
    \hline
    \multirow{2}{*}{\textbf{Frequency}} &
    \multirow{2}{*}{\textbf{\begin{tabular}[c]{@{}c@{}}Forecast \\ horizon\end{tabular}}}
    & \multicolumn{3}{c|}{\textbf{Test MAPE}}
    & \multicolumn{3}{c|}{\textbf{Average Runtime (minute)}} \\ \cline{3-8}
    &
    & \multicolumn{1}{l|}{\textbf{Silverkite}}
    & \multicolumn{1}{l|}{\textbf{FbProphet}}
    & \multicolumn{1}{l|}{\textbf{Auto-Arima}}
    & \multicolumn{1}{l|}{\textbf{Silverkite}}
    & \textbf{FbProphet}
    & \textbf{Auto-Arima} 
    \\ \hline
    \multirow{3}{*}{Daily}              & 1      &   40.97  &   64.78 & 41.83     & 0.6   & 2.2   & 0.4      \\ \cline{2-8}
                                        & 7      &   53.52  & 56.98   & 55.82     & 0.6   & 2.2   & 0.5      \\ \cline{2-8}
\hline
    \end{tabular}
    \caption{
        Summary of benchmarking results.
        MAPE and runtimes are averaged across multiple datasets.}
    \label{tab:benchmark-summary}
\end{table}

    \section{Discussion}
\label{sect:discussion}
This paper introduced a flexible framework for forecasting
which is designed for scalable and reliable forecasting
in production environments (Section \ref{sect:assessment}).

We showed how this design helps in generating flexible and interpretable forecasts
as well as volatility estimates.
As a particular example, the ability to use regularization algorithms as the
training algorithm of the mean component, allows us to accommodate complex patterns in the
model via feature interactions.

Having a separate volatility model, allows us to ensure fast speeds in production environments
where updating the forecast for many series might be needed.
However, this separation also helps in avoiding issues such as divergence of the simulated
series which are a common issue when using integrated models
(see \cite{hosseini-takemura-2015} and \cite{hosseini-bk-2020}).
The speed in fitting the model is also key in variable selection as many models can be fit to
optimize the choice of the component and parameters e.g.\;the Fourier series order for various time-scales
or the auto-regressive component complexity.

    \section{Appendix}
\label{sect:appendix}

\subsection{Solving the Mixed Regularization Problem}
\label{subsect:mix-regularization}
In this subsection,
we derive the two-step solution for the mixed penalty regression problem.
Without loss of generality,
we consider all weights equals 1, otherwise,
the same formulation can be obtained with a re-scale on the design matrix
and on the estimated coefficients.
Consider the regression problem
$$\hat{\bbeta}_0,\hat{\bbeta}_1,\hat{\bbeta}_2=\argmin_{\bbeta_0,\bbeta_1,\bbeta_2}\|\boldy - \boldX_0\bbeta_0 - \boldX_1\bbeta_1 - \boldX_2\bbeta_2\|_2^2 + \lambda_1\|\bbeta_1\|_1+\lambda_2\|\bbeta_2\|_2^2$$
Let,
$$\boldX_{02} = [\boldX_0,\boldX_2]$$
$$\bbeta_{02}=[\bbeta_0^T,\bbeta_2^T]^T$$
$$H_{02} = \boldX_{02}\left(\boldX_{02}^T\boldX_{02}\right)^{-1}\boldX_{02}^T$$
$$H_{\lambda02} = \boldX_{02}\left(\boldX_{02}^T\boldX_{02}+\lambda_2\boldD\right)^{-1}\boldX_{02}^T$$
where $\boldD$ is identity matrix with the first $m$ diagonal entries equal to zero,
and $m$ is the number of columns in $\boldX_0$. It's easy to verify that
$$H_{02}\boldX_0=\boldX_0$$
$$H_{02}\boldX_2=\boldX_2$$
We have:
\begin{align*}
    &\hat{\bbeta}_0,\hat{\bbeta}_1,\hat{\bbeta}_2\\
    =&\argmin_{\bbeta_0,\bbeta_1,\bbeta_2}\|\boldy - \boldX_0\bbeta_0 - \boldX_1\bbeta_1 - \boldX_2\bbeta_2\|_2^2 + \lambda_1\|\bbeta_1\|_1+\lambda_2\|\bbeta_2\|_2^2\\
    =&\argmin_{\bbeta_0,\bbeta_1,\bbeta_2}\|H_{02}(\boldy - \boldX_1\bbeta_1) - \boldX_0\bbeta_0 - \boldX_2\bbeta_2 + (\boldI-H_{02})(\boldy - \boldX_1\bbeta_1)\|_2^2 + \lambda_1\|\bbeta_1\|_1+\lambda_2\|\bbeta_2\|_2^2\\
    =&\argmin_{\bbeta_0,\bbeta_1,\bbeta_2}\|H_{02}(\boldy - \boldX_1\bbeta_1) - \boldX_{02}\bbeta_{02}\|_2^2 + \|(\boldI-H_{02})(\boldy - \boldX_1\bbeta_1)\|_2^2 + \lambda_1\|\bbeta_1\|_1+\lambda_2\|\bbeta_2\|_2^2\\
    =&\argmin_{\bbeta_1}\argmin_{\bbeta_{02}|\bbeta_1}\|H_{02}(\boldy - \boldX_1\bbeta_1) - \boldX_{02}\bbeta_{02}\|_2^2 + \|(\boldI-H_{02})(\boldy - \boldX_1\bbeta_1)\|_2^2 + \lambda_1\|\bbeta_1\|_1+\lambda_2\|\bbeta_2\|_2^2\\
\end{align*}
We have
\begin{align*}
    \hat{\bbeta}_{02}|\bbeta_1=&\argmin_{\bbeta_{02}|\bbeta_1}\|H_{02}(\boldy - \boldX_1\bbeta_1) - \boldX_{02}\bbeta_{02}\|_2^2 + \|(\boldI-H_{02})(\boldy - \boldX_1\bbeta_1)\|_2^2 + \lambda_1\|\bbeta_1\|_1+\lambda_2\|\bbeta_2\|_2^2\\
    =&\argmin_{\bbeta_{02}|\bbeta_1}\|H_{02}(\boldy - \boldX_1\bbeta_1) - \boldX_{02}\bbeta_{02}\|_2^2 + \lambda_2\|\bbeta_2\|_2^2\\
    =&\left(\boldX_{02}^T\boldX_{02}+\lambda_2\boldD\right)^{-1}\boldX_{02}^T(\boldy-\boldX_1\bbeta_1)
\end{align*}
Plugging back into the original equation, we have
\begin{align*}
    \hat{\bbeta}_1&=\argmin_{\bbeta_1}\|(H_{02}-H_{\lambda02})(\boldy-\boldX_1\bbeta_1)\|_2^2+\|(\boldI-H_{02})(\boldy - \boldX_1\bbeta_1)\|_2^2 + \lambda_1\|\bbeta_1\|_1+\lambda_2\|\bbeta_2\|_2^2\\
    &=\argmin\|(\boldI-H_{\lambda02})(\boldy - \boldX_1\bbeta_1)\|_2^2 + \lambda_1\|\bbeta_1\|_1\\
    &=\argmin\|(\boldI-H_{\lambda02})\boldy - (\boldI-H_{\lambda02})\boldX_1\bbeta_1\|_2^2+\lambda_1\|\bbeta_1\|_1
\end{align*}
This can be solved with the conventional Lasso algorithm with:
$$X=(\boldI-H_{\lambda02})\boldX_1$$
$$y=(\boldI-H_{\lambda02})\boldy$$

\end{document}